\documentclass[
 reprint,
 superscriptaddress,
 amsmath,amssymb,
 aps,
 prd,nofootinbib
]{revtex4-2}

\usepackage{graphicx}
\usepackage{dcolumn}
\usepackage{booktabs}
\usepackage[normalem]{ulem}
\usepackage{bm}

\usepackage{xcolor}

\usepackage{orcidlink} 

\usepackage{hyperref}

\begin{document}

\preprint{APS/123-QED}

\title{Exclusive Hadron Observables in Neutrino Induced $2p2h$ Multinucleon Knockout}

\author{Vedantha Srinivas Kasturi\,\orcidlink{0000-0001-7181-9150}}
\email{Vedantha.Kasturi@unige.ch}
\affiliation{Département de Physique Nucléaire et Corpusculaire, Université de Genève, Geneva, Switzerland}

\author{Juan Nieves\,\orcidlink{0000-0002-2518-4606}}
\affiliation{Instituto de F\'\i sica Corpuscular (IFIC), Centro Mixto CSIC-Universidad de Valencia, 
Institutos de Investigaci\'on de Paterna, Apartado 22085, E-46071 Valencia, Spain}

\author{Federico S\'anchez\,\orcidlink{0000-0003-0320-3623}}
\affiliation{Département de Physique Nucléaire et Corpusculaire, Université de Genève, Geneva, Switzerland}

\author{Joanna Ewa Sobczyk\,\orcidlink{0000-0003-4698-9339}}
\affiliation{Department of Physics, Chalmers University of Technology, Göteborg, Sweden}

\date{\today}

\begin{abstract}
We explore the combined lepton and hadron kinematic observables from the exclusive Valencia $2p2h$ model. 
We present variables of interest which are available due to the exclusive kinematics and compare them with the 
democratically distributed outgoing nucleon kinematics as currently treated in neutrino event generators. 
We also show the effect of nuclear re-scattering based on the NEUT semi classical cascade. 
We comment on the observability of these variables in current and future long baseline neutrino detectors. 
\end{abstract}

\maketitle


\section{\label{sec:intro}Introduction}

The importance of multinucleon mechanisms in neutrino-nucleus interactions has been established by several experiments in the last decade. 
The presence of these mechanisms was first proposed to explain the unexpectedly large charged current quasi elastic (CCQE) cross section measured by MiniBooNE \cite{MiniBooNE:2010bsu}. 
Since then, many theoretical models have been developed to describe these processes \cite{Martini:2009uj,Martini:2010ex,Nieves:2011pp,Martini:2011wp,Nieves:2011yp,Nieves:2012yz,Martini:2013sha,Nieves:2013fr,Megias:2014qva,Benhar:2015ula,Megias:2016fjk, Megias:2017cuh, Ivanov:2018nlm,Martinez-Consentino:2021vcs, Martinez-Consentino:2023dbt,Rocco:2026kae}.
These models have been successful in describing a variety of experimental data, including inclusive and semi-inclusive measurements from MiniBooNE 
\cite{MiniBooNE:2009dxl,MiniBooNE:2013qnd}, T2K \cite{T2K:2017rgv,T2K:2018rnz}, MINERvA \cite{MINERvA:2013kdn}, and NOvA \cite{NOvA:2021eqi}.

Multinucleon mechanisms, specifically two-particle-two-hole (2p2h) processes, involve the interaction of the neutrino with a correlated pair of nucleons  in the nucleus. 
This leads predominantly to the ejection of two nucleons from the nucleus along with the charged lepton, resulting in a final state with two holes in the nuclear ground state, in the mean-field picture. 
This process has been estimated to be ~20\% of the total CCQE cross section \cite{Martini:2009uj,Martinez-Consentino:2024mdr}, making it important in neutrino event generators that contain large fractions of CCQE-like events.
The kinematics of these processes are more complex than those of single nucleon knockout, as the energy and momentum transfer from the neutrino is shared between the two nucleons.
Additional complexity arises from the fact that the final state nucleons can undergo further interactions within the nucleus, leading to modifications of their kinematics and nucleon multiplicities that could mimic other interaction channels,
especially that of single nucleon knockout ($1p1h$) where only one nucleon is ejected. Having access to the hadronic kinematics is therefore crucial to disentangle these processes and further improve our understanding 
of neutrino-nucleus interactions.

The most common way to model outgoing kinematics of a $2p2h$ interaction in neutrino event generators is to assume that the two outgoing nucleons are emitted back-to-back in the center-of-mass (CM) frame of the interacting nucleon pair, and then transported through the nucleus. This simplified treatment does not take into account the internal dynamics of the process in the primary vertex of interaction. However, its simplicity allows to combine theoretical inclusive calculations of $2p2h$ with the intra-nuclear cascade.
This is the case of the charged-current model of Ref.~\cite{Nieves:2011pp}, which was implemented in most neutrino event generators, such as GENIE \cite{Andreopoulos:2009rq}, NEUT \cite{Hayato:2021heg}, and NuWro \cite{Prasad:2024gnv}. Since currently running long-baseline oscillation experiments like T2K rely on lepton kinematics alone to reconstruct the neutrino energy, this simplified treatment has been sufficient to describe the experimental data.

Current and next generation of neutrino detectors, such as T2K's ND280 upgrade \cite{T2K:2019bbb}, SBND \cite{MicroBooNE:2015bmn} and DUNE \cite{DUNE:2020lwj}, have the capability to measure exclusive hadronic final states with high precision. 
This opens up new opportunities to study multinucleon mechanisms in greater detail and to test the accuracy of the models used in neutrino event generators. In particular, the ability to measure the kinematics of the 
outgoing nucleons can provide valuable information about the underlying nuclear dynamics and the role of correlations in neutrino-nucleus interactions. This would only be possible with exclusive models capable 
of providing detailed information about the outgoing nucleons. This task is non-trivial due to the inherent complexity of the $2p2h$ response.
More degrees of freedom lead to a large dimensionality of an exclusive calculation, which makes it computationally expensive to implement in neutrino event generators.
One of the first models capable of providing exclusive kinematics for $2p2h$ interactions was developed in \cite{Sobczyk:2020dkn} built upon the inclusive  model of Ref.~ \cite{Nieves:2011pp}. Further refinements of the theoretical framework were presented in \cite{Sobczyk:2024ecl}, where the model of Ref.~\cite{Nieves:2011pp} was significantly improved. This updated framework (hereafter referred to as the Valencia model) forms the basis of our study. 

We would also like to mention the recent work of Ref.~\cite{Rocco:2026kae}, which extends the spectral-function formalism to exclusive two-nucleon knockout processes. This study introduces an improved treatment of the current operators entering the $\Delta$-current contribution, together with a more realistic description of the correlations between the two struck nucleons, including isospin dependence and angular correlations.

We explore the hadronic observables available from the exclusive Valencia $2p2h$ model and compare them with the simplified inclusive treatment currently used in neutrino event generators. Throughout this work, we focus on fixed neutrino energies of 450, 650 and 1000 MeV to isolate the model effects from the neutrino flux convolution and compare the two kinematic treatments across different variables of interest. In section~\ref{sec:exclusive_model}, we provide a brief overview 
of the exclusive Valencia $2p2h$ model. In section \ref{sec:inclusive_kinematics}, we describe the common inclusive approach used in neutrino event generators to compute the outgoing nucleon kinematics and how that varies from the exclusive model. Section \ref{subsec:nucleon_level} 
presents an exploration of the kinematic observables at the nucleon level and their theoretical relevance. In section \ref{subsec:fsi}, we study the effect of nuclear re-scattering (NrS) on these observables using the NEUT intranuclear cascade model. 
Finally, in section \ref{sec:observability}, we discuss the observability of these variables with detector limitations, and present our conclusions in section \ref{sec:concl}
\section{Exclusive Valencia $2p2h$ model}
\label{sec:exclusive_model}

The Valencia model for $2p2h$ interactions was originally developed to describe inclusive photon~\cite{Carrasco:1989vq}, electron~\cite{Gil:1997bm}, and pion scattering from nuclei~\cite{Salcedo:1987md,Garcia-Recio:1989hyf,Nieves:1993ev,Nieves:1991ye}, where multi-nucleon mechanisms play a crucial role. It was first applied to charged-current neutrino–nucleus interactions in Ref.~\cite{Nieves:2011pp}, in conjunction with the $1p1h$ contribution (quasi-elastic (QE)) derived in Ref.~\cite{Nieves:2004wx}, both based on the same microscopic many-body framework. More recently, the model of Ref.~\cite{Nieves:2011pp} was updated in Ref.~\cite{Sobczyk:2024ecl}, as mentioned above.

The scheme is based on a local Fermi gas description of the nucleus and starts from a state-of-the-art description of (anti-)neutrino-induced pion production off free nucleons~\cite{Hernandez:2007qq,Hernandez:2010bx,Alvarez-Ruso:2015eva,Hernandez:2016yfb}. It includes contributions from both meson-exchange currents and final-state nucleon correlations, while only statistical correlations are included in the initial state, in accordance with the established power-counting scheme. An additional key ingredient of the approach is the effective in-medium baryon--baryon interactions in the $NN$, $N\Delta$, and $\Delta\Delta$ channels, which have been successfully tested in a wide range of reactions at intermediate energy transfers. A distinctive theoretical feature of this approach, compared to other $2p2h$ models used in neutrino--nucleus interactions, is that it goes beyond the one-pion-exchange potential. Both spin longitudinal and transverse channels are considered, driven by $\pi$ and $\rho$ exchange, respectively, with short-range correlations accounted for by contact terms in the interaction.
 
The calculations of Ref.~\cite{Nieves:2011pp} (or Ref.~\cite{Sobczyk:2024ecl}) provided the inclusive cross sections, where only the outgoing lepton is detected.
However, recent works have extended the approach to allow for the computation of exclusive kinematics. This was achieved by removing some of the numerical approximations and, most importantly, by a direct calculation of the $\Delta\Delta$ contribution~\cite{Sobczyk:2020dkn}. In particular, the $\Delta$ in-medium treatment plays a crucial role in the dynamics of $2p2h$ process. In the most recent calculation presented in Ref.~\cite{Sobczyk:2024ecl}, the effective real part of the $\Delta$ self-energy was used to derive a robust estimate of the theoretical uncertainty. This is because, depending on whether the $\Delta$ is excited by a longitudinal spin operator, as in the case of pion-nucleus scattering, or by a transverse one, as in the case of photon-nucleus scattering, the in-medium shift of the resonance position takes very different values. In the case of neutrinos, both types of spin operators appear in the excitation of the $\Delta$.  In Ref.~\cite{Sobczyk:2024ecl}, a comprehensive discussion of the $2p2h$ contributions is presented within a many-body quantum theory framework, and several improvements over the 2011 calculation of Ref.~\cite{Nieves:2011pp} are implemented. In particular, only spin-3/2 degrees of freedom are retained in the $\Delta$ propagators, which significantly improves the description of the $\nu_\mu n \to \mu^- n \pi^+$ reaction and, more generally, leads to a better fulfillment of Watson’s theorem~\cite{Hernandez:2016yfb}. In addition, an earlier inconsistency in the treatment of in-medium nucleon energy, was resolved in~\cite{Sobczyk:2024ecl}, leading to larger strength of the $2p2h$ nuclear-response and a better agreement with the T2K and MiniBooNE data, as well as theoretical calculation by Martini et al.~\cite{Martini:2009uj}. 

\begin{figure}[htbp]
    \centering
    \includegraphics[width=\linewidth, height=0.5\linewidth]{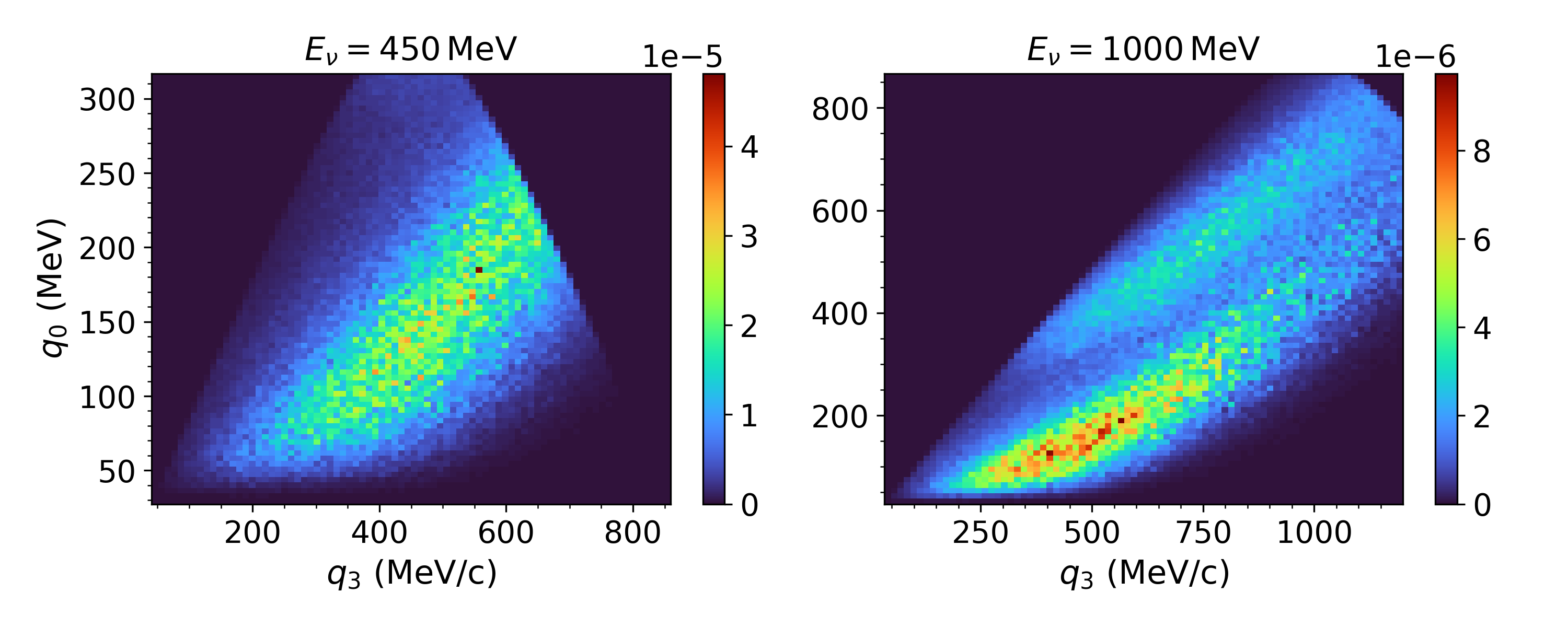}
    \caption{Cross section weighted 2D distribution in energy $q_0$ (y axis) and momentum transfer $|\vec q\,|$ (x axis) plane for Valencia $2p2h$ model at increasing neutrino energies. Colours indicates the unit-normalised 2D density. We observe that the $\Delta$-resonance peak becomes visible at higher energy transfers for the highest neutrino energy.}
    \label{fig:Valencia_2p2h_CrossSection_q0_q3}
\end{figure}

\section{Inclusive Hadron kinematics computation}
\label{sec:inclusive_kinematics}


Current implementations of the Valencia $2p2h$ model in neutrino event generators use hadron tensor tables to compute the inclusive cross sections obtained in Ref.~\cite{Nieves:2011pp}.
Hadron tensor tables are multidimensional arrays that compute the hadronic tensor values as functions of energy-momentum transfer, $q^\mu=(q_0, \vec{q}\,)$, allowing event generators to efficiently compute 
cross sections without re-evaluating them for each event. This method was first used in the implementation of the Valencia $2p2h$ model in NEUT and was later adopted in other generators \cite{Schwehr:2016pvn}. For a given nucleus, the tables are computed in bins of $q_0$ and $|\vec{q}\,|$, energy and momentum transferred to the nuclear system, respectively. 
This approach then allows for fast computation of the cross sections during event generation by interpolating the hadron tensor values from the tables based on the kinematics of the incoming neutrino and outgoing lepton.

The choice of variables is motivated by their initial use in inclusive models and the need to keep these tables manageable in size and computationally efficient. The kinematics of the hadronic system are then approximated respecting energy and momentum conservation.
Since no information about the initial nucleon momenta is retained, they are randomly sampled and simplified assumptions are made about the kinematics of the outgoing nucleons based on available energy and momentum transfer.

The computation of the outgoing nucleon kinematics in the inclusive approach typically involves the following steps. The initial nucleons' momenta $p_{N_i}$ where $i = 1,2$ are sampled from a local Fermi gas distribution. The initial nucleons are treated as on-shell particles
\begin{equation}
    E_{N_i} = \sqrt{\vec{p}_{N_i}^{\,2} + m_N^2}\, , \quad
    \vec{p}_{N_i} = |\vec{p}_{N_i}|\,\hat{u}_i\, ,
\end{equation}
where $\hat{u}_i$ is a random unit vector representing the direction of the nucleon momentum. The final-state four-momentum (accounting for the separation energy of a neutron-proton pair which for $^{12}$C is\footnote{For a proton-proton pair, the separation energy is around 0.2 MeV smaller.} $E_S = 27.41$ MeV) is then given by:
\begin{equation}
    P_{\mathrm{final}}^\mu = P_{N_1}^\mu + P_{N_2}^\mu + q^\mu - (E_S, \vec{0})\, .
\end{equation}
with $q^\mu$ Monte Carlo distributed according the updated theoretical $d^2\sigma/(dq^0d|\vec{q}\,|)$ differential cross section calculated in ~\cite{Sobczyk:2024ecl} and shown here in Fig.~ \ref{fig:Valencia_2p2h_CrossSection_q0_q3}. At this stage, one has to ensure that sufficient energy is available to eject the two nucleons from the nucleus. This is done by requiring that the invariant mass $\sqrt{S} = \sqrt{{P_{\mathrm{final}}^\mu}^2} > 2 m_N$. Initial nucleons are sampled until this condition is satisfied.
Once this is done, the outgoing nucleon four-momenta in the two-nucleon  CM frame are given by:
\begin{equation}
\begin{aligned}
E^{\mathrm{cm}} &= \frac{\sqrt{S}}{2}, \\
|\vec{p}^{\,\mathrm{cm}}| &= \sqrt{\left(E^{\mathrm{cm}}\right)^2 - m_N^2}, \\
\vec{p}^{\,\mathrm{cm}}_{1} &= |\vec{p}^{\,\mathrm{cm}}|\, \hat{r}, \qquad
\vec{p}^{\,\mathrm{cm}}_{2} = -\vec{p}^{\,\mathrm{cm}}_{1}, \\
P^{\mu\,\mathrm{cm}}_{1,2} &= \bigl(E^{\mathrm{cm}},\, \vec{p}^{\,\mathrm{cm}}_{1,2}\bigr).
\end{aligned}
\end{equation}
where $\hat{r}$ is a random unit vector representing the direction of $\vec{p}^{\,\mathrm{cm}}_{1}$. Finally, the outgoing nucleon four-momenta are boosted back to the laboratory frame using:
\begin{equation}
\vec{\beta} = \frac{\vec{P}_{\mathrm{final}}}{P^{0}_{\mathrm{final}}}\, , \quad
P'_{1,2} \equiv P^{\mathrm{lab}}_{1,2} = \Lambda(\vec{\beta}) \cdot P^{\mathrm{cm}}_{1,2}\, ,
\end{equation}
where $\Lambda(\vec{\beta})$ is the Lorentz boost operator.
Approximations made here result in identical behavior for both outgoing nucleons with regards to their correlations with the lepton kinematics. Any asymmetry beyond that arising from the NrS is not a result of this treatment. In effect, one is blind to the internal dynamics of the nucleon pair. Another important aspect to note is the loss of information about the exact initial state correlations between the nucleons. Hence, this approach has several disadvantages: it affects the observable quantities, limit the ability to study the underlying nuclear dynamics in detail and restrict model discrimination to total cross sections and lepton kinematics alone. In order to understand the impact of these simplifications, we compare the hadronic observables obtained using the inclusive approach with those obtained using the exclusive kinematics from the Valencia model. To isolate the effects of the kinematic treatments from those arising from cross‑section computational differences \cite{Sobczyk:2024ecl}, and to focus on observable outgoing nucleon dependent quantities, we rely on the same underlying theoretical framework, namely Refs.~\cite{Sobczyk:2020dkn,Sobczyk:2024ecl}, consistently adopting the improvements introduced in \cite{Sobczyk:2024ecl} into the original $2p2h$ model of Ref.~\cite{Nieves:2011pp}.

\begin{figure}[htbp]
    \centering
    \includegraphics[width=0.494\linewidth, height=0.44\linewidth]{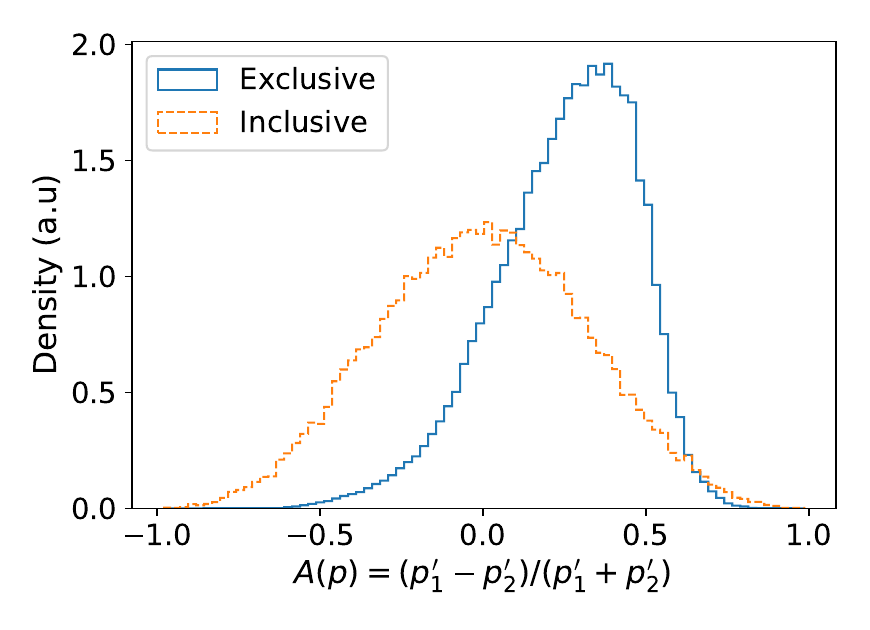}
    \includegraphics[width=0.494\linewidth, height=0.44\linewidth]{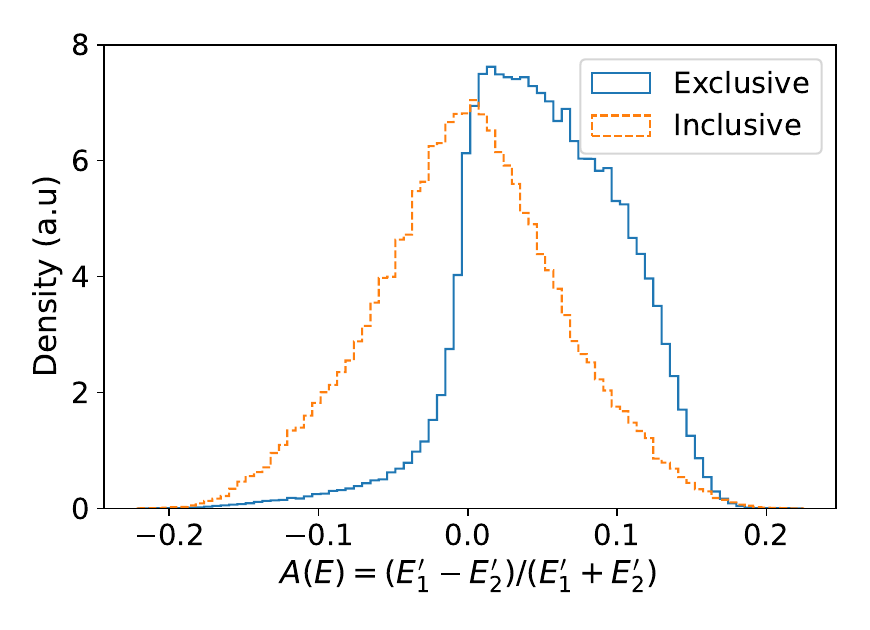}
    \caption{Normalised event distributions (PDFs) of the asymmetries in Eq.~\eqref{eq:Asymmetry} for the $pp$ final-pair channel at $E_\nu=650~\mathrm{MeV}$. Left: momentum asymmetry $A(p)$; right: energy asymmetry $A(E)$.}
    \label{fig:Asymmetry_inclusive_vs_Exclusive}
\end{figure}

We quantify the event-by-event imbalance between the two outgoing nucleons using the
momentum and energy asymmetries,
\begin{equation}
A(p) = \frac{p_1^{\prime} - p_2^{\prime}}{p_1^{\prime} + p_2^{\prime}}, \qquad
A(E) = \frac{E_1^{\prime} - E_2^{\prime}}{E_1^{\prime} + E_2^{\prime}},
\label{eq:Asymmetry}
\end{equation} 
where $p_i^{\prime} \equiv |\vec{p}_{N_i^{\prime}}|$ and $E_i^{\prime} \equiv E_{N_i^{\prime}}$ ($i=1,2$) are, respectively, the magnitude of the three-momentum and the total energy of the two outgoing nucleons in the lab frame.\footnote{Primes denote final-state (outgoing) nucleon kinematics.} The interpretation of the labels $i=1,2$ differs between the inclusive and exclusive calculations. In the inclusive model, the two outgoing nucleons are treated symmetrically by construction (see Eqs.~(1)--(4)), so the assignment ``1'' versus ``2'' carries no physical meaning; consequently, $A(p)$ and $A(E)$ are expected to be symmetric around zero when evaluated over an event sample. In the exclusive model, instead, the two nucleons are distinguished by the interaction mechanism: one nucleon is directly associated with the $W$ vertex—that is, it absorbs the effective energy transfer $(q^0 - E_S)$ and the full momentum transfer $\vec{q}$—while the other is the accompanying nucleon. This physically motivated labeling breaks the $1\leftrightarrow 2$ symmetry and can produce asymmetric $A(p)$ and $A(E)$ distributions. Fig~\ref{fig:Asymmetry_inclusive_vs_Exclusive} shows these distributions for the mono-energetic sample at $E_\nu = 650~\mathrm{MeV}$ in the $pp$ final-pair channel, where the exclusive treatment shows a distinctive asymmetry.
This highlights the limitation of the inclusive approximation, which assumes symmetric momentum sharing and fails to capture the complex nucleon dynamics in $2p2h$ interactions. It is important to note that this plot shows nucleon level kinematics still inside nuclear medium and does not include final state interactions (FSI). The latter effect, specifically NrS will be discussed in Section \ref{subsec:fsi}.
\section{Exploration of Exclusive Kinematics}
\label{sec:exploration}
Historically, neutrino scattering experiments have had limited capability to measure exclusive hadronic final states with sufficient precision to directly constrain the models implemented in event generators. 
This situation is evolving rapidly with upcoming high-statistics experiments and major detector upgrades that offer significantly enhanced hadronic reconstruction capabilities.
It is therefore timely to identify which hadronic observables are both sensitive to the choice of kinematic treatment (exclusive versus inclusive) and experimentally accessible. 
In this section, we systematically explore the  exclusive Valencia $2p2h$ kinematics derived from the updated model of Ref.~\cite{Sobczyk:2024ecl}, and identify key variables that can discriminate between kinematic prescriptions in current and future neutrino detectors.

\subsection{Nucleon-level observables}
\label{subsec:nucleon_level}
At the primary interaction vertex, the process is modeled within an impulse approximation where the neutrino interacts with a correlated pair of nucleons in the nucleus. The relevant Feynman diagrams for this process
can be seen in Fig.~1 of Ref.~\cite{Sobczyk:2024ecl}.
The nuclear dynamics of the process influences how the transferred energy is shared between the two interacting nucleons. The kinematics of the outgoing nucleons can be determined from energy and momentum conservation,
\begin{equation}
\begin{aligned}
E_\nu + E_{N_1} + E_{N_2} &= E_\ell + E_{N_1'} + E_{N_2'} - E_S, \\
\vec{p}_\nu + \vec{p}_{N_1} + \vec{p}_{N_2} &= \vec{p}_\ell + \vec{p}_{N_1'} + \vec{p}_{N_2'}.
\end{aligned}
\end{equation}
where $E_\nu$ and $\vec{p}_\nu$ are the energy and momentum of the incoming neutrino, $E_\ell$ and $\vec{p}_\ell$ are the energy and momentum of the outgoing lepton,
$E_{N_1}$ and $\vec{p}_{N_1}$  and $E_{N_2}$ and $\vec{p}_{N_2}$  are the energy and momentum of the pair of initial nucleons, while  $E_{N_1'}$ and $\vec{p}_{N_1'}$  and $E_{N_2'}$ and $\vec{p}_{N_2'}$ are the energy and momentum of the pair of outgoing nucleons. 
This now provides us with the full kinematics of all 6 particles in the initial and final states: 
the initial on-shell nucleon momenta are randomly chosen at point $\vec{r}$ of the nucleus and given these four-momenta, $q^0$ and $\vec{q}$ and the outgoing nucleon momenta are Monte Carlo generated according to the theoretical differential exclusive distribution $d\sigma/(d^3r\, d^3p_{N_1}d^3p_{N_2} d^3p_{N_1'}d^3p_{N_2'} d^4q)$ as calculated in \cite{Sobczyk:2020dkn}, but now using the updated $2p2h$ nuclear model of   Ref.~\cite{Sobczyk:2024ecl}. 

 In practice, the observation depends on the detector capabilities to identify neutrons and protons, which is more likely for the nucleon with higher momenta. Therefore all the comparisons between nucleons will focus on leading and sub-leading nucleons, defined as the nucleon with the highest and second highest momentum respectively. For the charged-current $2p2h$ reaction we distinguish three possible configurations for the outgoing nucleons for the neutrino: proton-proton [$pp$], proton-neutron [$pn$] and neutron-proton [$np$] (for the antineutrino these are neutron-neutron [$nn$], neutron-proton [$np$] and proton-neutron [$pn$]). Note that $nn$ ($pp$) pairs can be produced in neutrino (antineutrino) reactions only thanks to secondary FSI.

As mentioned, the model distinguishes between the nucleons carrying the highest and second-highest momenta, which may be associated, respectively, with the nucleon that interacts with the $W$ boson and with the nucleon that participates through its interaction with the former, as introduced in the previous discussion of Fig.~\ref{fig:Asymmetry_inclusive_vs_Exclusive}. Hence, it is useful to distinguish between $pn$ and $np$ final states at the level of the model. It should be noted, however, that experimentally both correspond to the same physical final state, namely a neutron–proton pair. 

To study the region of interest for $2p2h$, we will focus on the neutrino induced states and at fixed neutrino energies of 450, 650 and 1000 MeV, with earlier emphasis on 650 MeV due to its proximity to the T2K and Hyper-Kamiokande's flux peak \cite{Hyper-Kamiokande:2018ofw, T2K:2019bbb}.
The relative fraction of the total neutrino $2p2h$ cross section for each channel is shown in Table \ref{tab:channel-fractions}. Since the majority (80\%) of the cross-section comes from the $pp$ channel, we will primarily focus on this for the nucleon level observables, though we will also examine combined ($pp+pn$) distributions where relevant. 
\begin{table}[htbp]
    \centering
    \begin{tabular}{l c c}
        \toprule
        Neutrino energy (MeV) & $pp$ (\%) & $pn{+}np$ (\%) \\
        \midrule
        450  & 80.3 & 19.7 \\
        650  & 80.5 & 19.5 \\
        1000 & 80.5 & 19.5 \\
        \bottomrule
    \end{tabular}
    \caption{Fraction of total neutrino $2p2h$ cross section by outgoing channel after the primary interaction with the $W$ boson. Values are identical for inclusive and exclusive treatments since they are based on the same underlying model.}
    \label{tab:channel-fractions}
\end{table}
\subsubsection{Lepton and hadron correlations}
\label{subsubsec:lepton_hadron}
Oscillation experiments typically rely on outgoing lepton kinematics, since the lepton is the most reliably reconstructed final-state particle. However, the hadronic system carries a significant fraction of the energy–momentum, and lepton–hadron correlations encode nuclear effects that can bias the reconstruction. We therefore study correlations between the outgoing nucleons and their relation to the outgoing lepton. We focus on the nucleon momenta and the lepton polar angle with respect to the incoming neutrino.
\begin{figure}[htbp]
    \centering
    \includegraphics[width=\linewidth, height=0.5\linewidth]{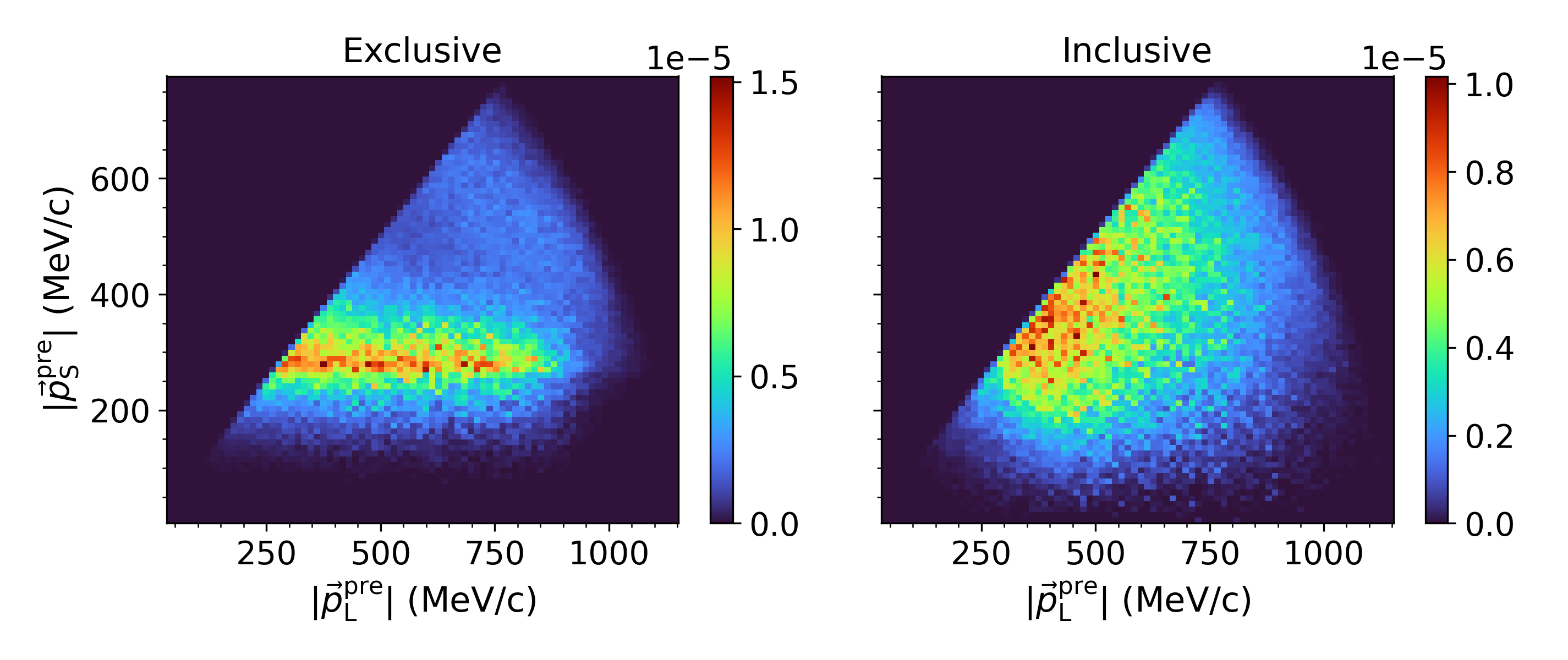}
    \caption{Cross-section--weighted 2D PDF of $|\vec{p}_{\mathrm{L}}^{\,\mathrm{pre}}|$ (x-axis) vs $|\vec{p}_{\mathrm{S}}^{\,\mathrm{pre}}|$ (y-axis) for $E_\nu=650~\mathrm{MeV}$ in the $pp$ channel. Colour indicates the unit-normalised 2D density. Exclusive (left) vs inclusive (right).}
    \label{fig:NucleonMomentum_LeptonAngle_Correlation}
\end{figure}    
Fig. \ref{fig:NucleonMomentum_LeptonAngle_Correlation} shows the primary difference between the inclusive and exclusive treatments. Secondary collisions are not taken into account. In the inclusive treatment, the nucleon momenta are more symmetrically distributed, losing information from the vertex dynamics. In the exclusive calculation, we see a clear asymmetry between the leading $|\vec{p}_L^{\rm \,pre}|$ and subleading nucleon $|\vec{p}_S ^{\rm \,pre}|$ momenta, with the leading nucleon typically carrying a larger fraction of the momentum transfer \footnote{$pre$ and $post$ refer to quantities before and after NrS}. This is consistent with the expectation that one nucleon interacts directly with the $W$ boson, receiving a larger share of the energy-momentum transfer, while the second nucleon receives less momentum via meson exchange.

\begin{figure}[htbp]
    \centering
    \includegraphics[width=\linewidth,height=0.9\linewidth]{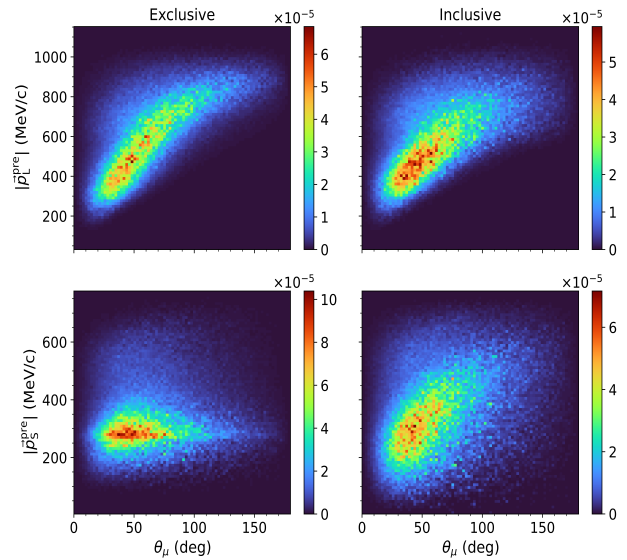}
    \caption{Cross-section--weighted 2D PDF of $|\vec{p}_{\mathrm{L}}^{\,\mathrm{pre}}|$ (top) \& $|\vec{p}_{\mathrm{S}}^{\,\mathrm{pre}}|$ (bottom) (y-axis) vs $\theta_\mu$ (x-axis) for $E_\nu=650~\mathrm{MeV}$ in the $pp$ channel. Colour indicates the unit-normalised 2D density. Exclusive (left) vs inclusive (right).}
    \label{fig:NucleonMomentum_LeptonAngle_CorrelatioN_2}
\end{figure}

The energy transfer to the hadronic system is highly correlated to the lepton angle. Hence we expect a correlation between the lepton angle $\theta_\mu$ and the outgoing nucleon momenta. This, however, is complicated in the $2p2h$ case due to the meson exchange mechanism that splits the total energy-momentum transfer between the two nucleons. We first pay attention to the leading nucleon, and in  Fig.~\ref{fig:NucleonMomentum_LeptonAngle_CorrelatioN_2},  we observe this correlated behavior in both treatments, with a stronger relation seen in the exclusive one. The same cannot be said for the comparisons with the sub-leading nucleon. The inclusive treatment results in a distribution similar to that obtained for the leading nucleon, while the exclusive treatment shows a subleading nucleon that is largely independent of the lepton kinematics. In the exclusive treatment, most of the energy--momentum transfer is carried by the nucleon attached to the primary $W$ vertex. As a result, the momentum of this nucleon (which is most often the leading proton) closely tracks the lepton kinematics and produces the stronger correlation. The second nucleon, associated with the meson-exchange vertex, typically receives a smaller share of the transfer and is therefore much less correlated with the lepton angle, populating predominantly lower momenta and a broader angular phase space as compared to the inclusive case.
These differences shown in Fig.~\ref{fig:NucleonMomentum_LeptonAngle_Correlation},~\ref{fig:NucleonMomentum_LeptonAngle_CorrelatioN_2} hold across the three neutrino energies tested. In exclusive treatment the leading nucleon carrying the major fraction of the transfer momenta and the sub-leading nucleon largely flat with respect to lepton angle while the inclusive behavior is same across the board by construction.
\subsubsection{Transverse kinematic imbalance}
\label{subsubsec:tki}
\begin{figure*}[htbp]
    \centering
    \includegraphics[width=\textwidth]{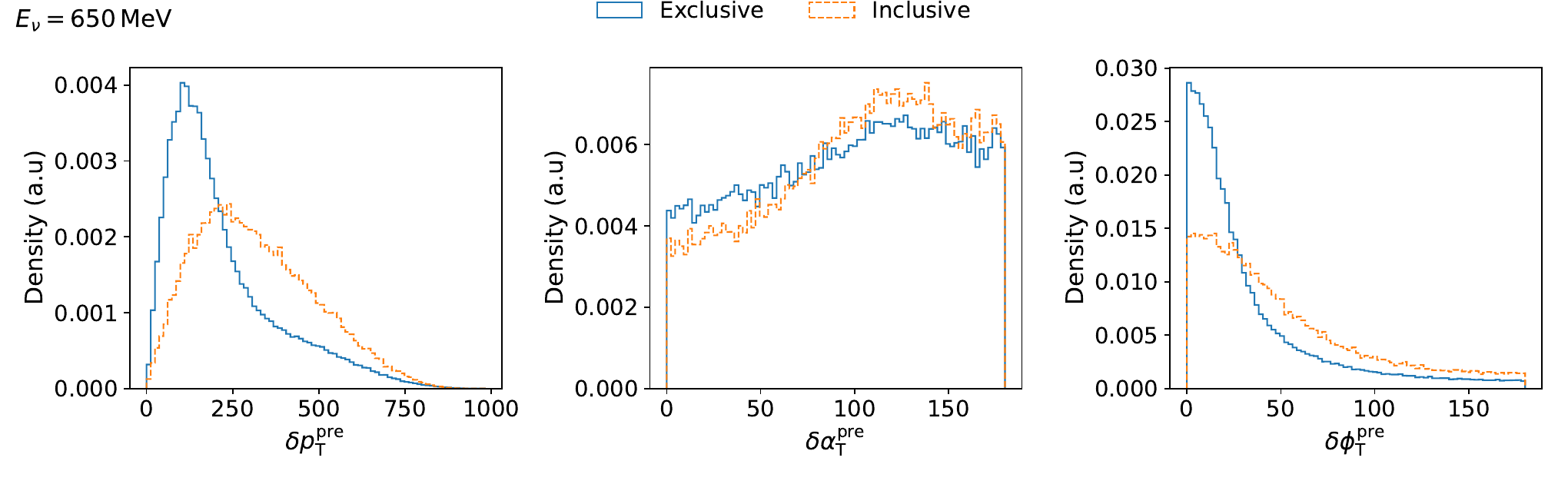}
    \caption{TKI variables (inclusive versus exclusive) for $E_\nu=650$ MeV $(pp+pn)$. Left, middle and right top panels show the obtained distributions for  $\delta p_T$, $\delta \alpha_T$ and $\delta \phi_T$, respectively.}
    \label{fig:tKI_Variables_PreFsi}
\end{figure*}

The importance of transverse kinematic imbalance (TKI) variables in neutrino-nucleus interactions has been increasingly recognized in recent years \cite{Lu:2015tcr,MINERvA:2018hqn,Bourguille:2020bvw}.
These variables are constructed from the transverse components of the momenta of the outgoing particles in the final state, and they provide valuable information about the nuclear effects and the underlying interaction mechanisms. 
In particular, TKI variables are sensitive to multinucleon effects, such as $2p2h$ interactions especially if they are reconstructed with the assumption of a single outgoing nucleon observed. In our study
we pick the leading nucleon as the observed nucleon. The TKI variables we consider are the transverse momentum imbalance $\delta p_T$, the transverse opening angle imbalance $\delta \alpha_T$, and the angular transverse imbalance $\delta \phi_T$ given by:
\begin{equation}
    \delta p_T = |\vec{p}_T^{\, \ell} + \vec{p}_T^{\,N_{\rm lead}}|
\end{equation}
\begin{equation}
    \delta \alpha_T = \arccos\left(\frac{-\vec{p}_T^{\, \ell} \cdot \delta \vec{p}_T}{|\vec{p}_T^{\, \ell}||\delta \vec{p}_T|}\right)
\end{equation}
\begin{equation}
    \delta \phi_T = \arccos\left(\frac{-\vec{p}_T^{\, \ell} \cdot \vec{p}_T^{\, N_{\rm lead}}}{|\vec{p}_T^{\, \ell}||\vec{p}_T^{\,N_{\rm lead}}|}\right)
\end{equation}  
where $\vec{p}_T^{\, \ell}$ and  $\vec{p}_T^{\, N_{\rm lead}}$ are the the outgoing lepton and leading nucleon transverse momenta,  and $\delta \vec{p}_T$ is the transverse momentum imbalance vector.~\footnote{Note that $\vec{p}^{ N_{\rm lead}}$ was denoted as $\vec{p}_{\mathrm{L}}^{\,\mathrm{pre}}$ in the previous section. We have changed the notation here to avoid confusion with the $T$ subscript used to denote the transverse component.}
Fig.~\ref{fig:tKI_Variables_PreFsi} shows the distributions of these variables in the exclusive treatment and compares them with those obtained in the inclusive treatment.
Since these variables are constructed from an assumed observation of single nucleon, they are highly sensitive to the outgoing nucleon configurations and the neutrino energy. 
Hence we explore the combined ($pp+pn$) distributions. This means that the interaction with the $W$ boson occurs on $pn$ or $nn$ pairs. We emphasize that, up to this stage, secondary collisions have not yet been taken into account. These effects will be addressed in the following subsection. As can be observed in Fig.~\ref{fig:tKI_Variables_PreFsi}, the differences between the inclusive and exclusive treatments 
are clearly visible. The favoring of leading nucleon in the exclusive treatment leads to a more QE-like behavior. The variations across $E_\nu$ are shown in Sec \ref{subsec:nd280}.

This behavior is clearly visible in the $\delta p_T$ distribution, where the exclusive treatment exhibits a pronounced peak at low transverse momentum imbalance, whereas the inclusive treatment yields a broader distribution extending to higher $\delta p_T$ values. The physical origin is straightforward: when the leading nucleon carries most of the transferred momentum (as in the exclusive calculation) and the lepton kinematics are held fixed, the vector sum $\vec{p}_T^{\, \ell} + \vec{p}_T^{\, N_{\rm lead}}$ tends toward cancellation, resulting in small $\delta p_T$ and thus a QE-like signature. Conversely, the inclusive prescription, which distributes energy and momentum more symmetrically between the two outgoing nucleons, produces a less complete cancellation between lepton and leading-nucleon transverse momenta, thereby shifting the $\delta p_T$ distribution to higher values.


The $\delta \alpha_T$ distribution shows less variation between the two treatments but nonetheless exhibits a non-flat shape across the angles in the inclusive case. This arises due to the approximated angular distribution of the nucleon being insufficient to balance with the muon, inevitably creating angular momentum imbalance in the system. The angular sensitivity becomes more pronounced at higher neutrino energies, where the kinematic differences between treatments for the missing nucleon are larger when compared to the exclusive case where most of it goes to the leading nucleon with a more accurate angular distribution and tending towards a flatter profile.

The transverse azimuthal angle difference $\delta \phi_T$ exhibits a distribution characteristic of QE scattering when analyzed using the exclusive event selection. Specifically, the distribution demonstrates a pronounced maximum at $\delta \phi_T \approx 0°$, indicating back-to-back momentum topology in the transverse plane. This collimation is once again the result of the leading nucleon preference within the exclusive treatment, which effectively mimics QE-like kinematics. In contrast, the inclusive treatment yields a less peaked and flatter $\delta \phi_T$ distribution, reflecting the more symmetric momentum sharing between the two outgoing nucleons. The peak nonetheless arises from picking the leading nucleon alone, which leaves an imbalance that is still present.
One essential point to note is that these are computed for all outgoing leading nucleons detected irrespective of isospin. This is not always the case in experiments where the proton detection is much more efficient than neutron detection. Another important aspect is the treatment of NrS, which can affect the observables significantly by smearing out the variations.
\subsection{Effect of nuclear rescattering}
\label{subsec:fsi}
Nuclear re‑scattering in event generators is typically treated with semi‑classical hadronic intranuclear cascade  simulations. These algorithms propagate the primary outgoing hadrons through the nuclear medium and account for processes such as elastic and inelastic scattering, charge‑exchange, absorption and secondary nucleon knockout, together with effects like Pauli blocking and momentum loss. The cascade evolution is governed by the hadron–nucleon interaction cross sections (microscopic or parameterized) together with the local nuclear density, which sets the mean free path. In this work both the exclusive and inclusive event samples are passed through the NEUT intranuclear cascade to quantify the impact of NrS on the exclusive kinematics and on the observable distributions \cite{Hayato:2021heg}.

\begin{figure}[htbp]
    \centering
    \includegraphics[width=\linewidth, height=0.5\linewidth]{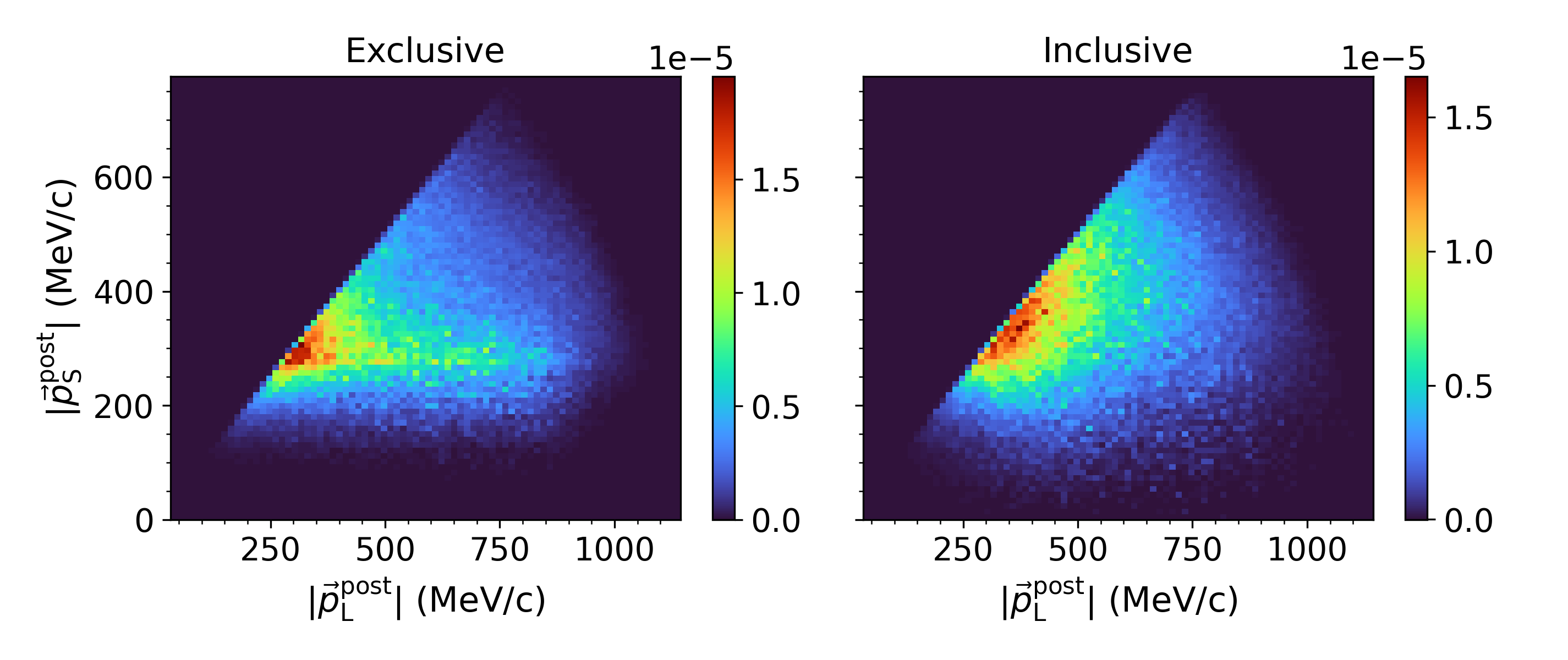}
    \caption{Same as in Fig.~\ref{fig:NucleonMomentum_LeptonAngle_Correlation}, but considering NrS $(pp+pn)$.} 
    \label{fig:NucleonMomentum_Correlation_FSI}
\end{figure}

The number of outgoing nucleons is not conserved after NrS. Primary nucleons can be absorbed while secondary nucleons may be knocked out; processes such as charge exchange or pion production can modify the final‑state isospin composition. For the observables in this study we therefore select the two highest‑energy nucleons in the final state. This choice is motivated by two considerations: (i) at the neutrino energies considered, the probability of emitting more than two energetic nucleons is small, so the two leading nucleons dominate the experimentally detectable signal, and (ii) using the leading pair provides a consistent, experimentally relevant definition that enables a fair comparison between the inclusive and exclusive kinematic treatments and to understand the impact of NrS on these observables.

\begin{figure}[htbp]
    \centering
    \includegraphics[width=\linewidth, height=0.9\linewidth]{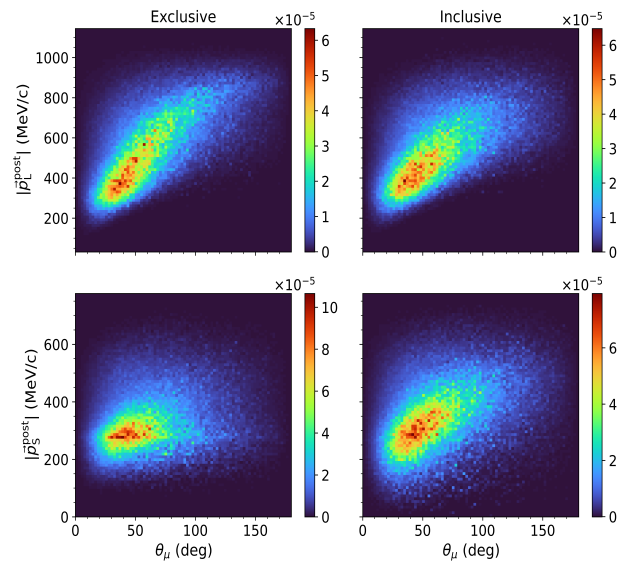}
    \caption{Same as Fig.~\ref{fig:NucleonMomentum_LeptonAngle_CorrelatioN_2}, but considering NrS $(pp+pn)$.}
    \label{fig:LeptonAngle_NucleonMomentum_Correlation_FSI}
\end{figure}

NrS primarily broadens and redistributes the outgoing nucleon kinematics: additional elastic/inelastic scatterings, charge exchange and secondary knockout soften the spectra and wash out sharp structures.  Despite this NrS-induced smearing, the exclusive kinematic calculation continues to predict a higher‑momentum leading nucleon than the inclusive prescription.  This persistence indicates that the tendency for a single nucleon to carry a dominant fraction of the transferred energy–momentum is a genuine feature of the exclusive dynamics and is not erased by the cascade. Correspondingly, the exclusive model yields a softer subleading‑nucleon spectrum compared with the inclusive treatment.  Reassessing the correlations discussed in Sec.~\ref{subsec:nucleon_level}, Fig.~\ref{fig:NucleonMomentum_Correlation_FSI} shows that the post‑NrS leading versus subleading momentum correlation still exhibits the stronger asymmetry characteristic of the exclusive calculation.


Examination of the lepton‑angle versus subleading‑nucleon momentum correlation in Fig.~\ref{fig:LeptonAngle_NucleonMomentum_Correlation_FSI} reveals that the exclusive calculation  shows a significantly weaker correlation with lepton angle, especially at higher momenta, whereas the inclusive prescription preserves a residual dependence on the lepton polar angle $\theta_\mu$ similar to what was observed pre NrS. For the leading nucleon (highest‑energy) distribution, both treatments exhibit a correlation with the lepton angle $\theta_\mu$, but the exclusive model shows higher concentration of harder leading‑nucleon spectrum at large $\theta_\mu$ than the inclusive model. This difference is experimentally relevant: detectors with limited acceptance at large lepton scattering angles will sample these correlated lepton-nucleon phase‑space regions differently, potentially producing model‑dependent detection rates for events with both a muon and proton and affecting the sensitivity to multinucleon dynamics.

\begin{figure*}[htbp]
    \centering
    \includegraphics[width=\linewidth]{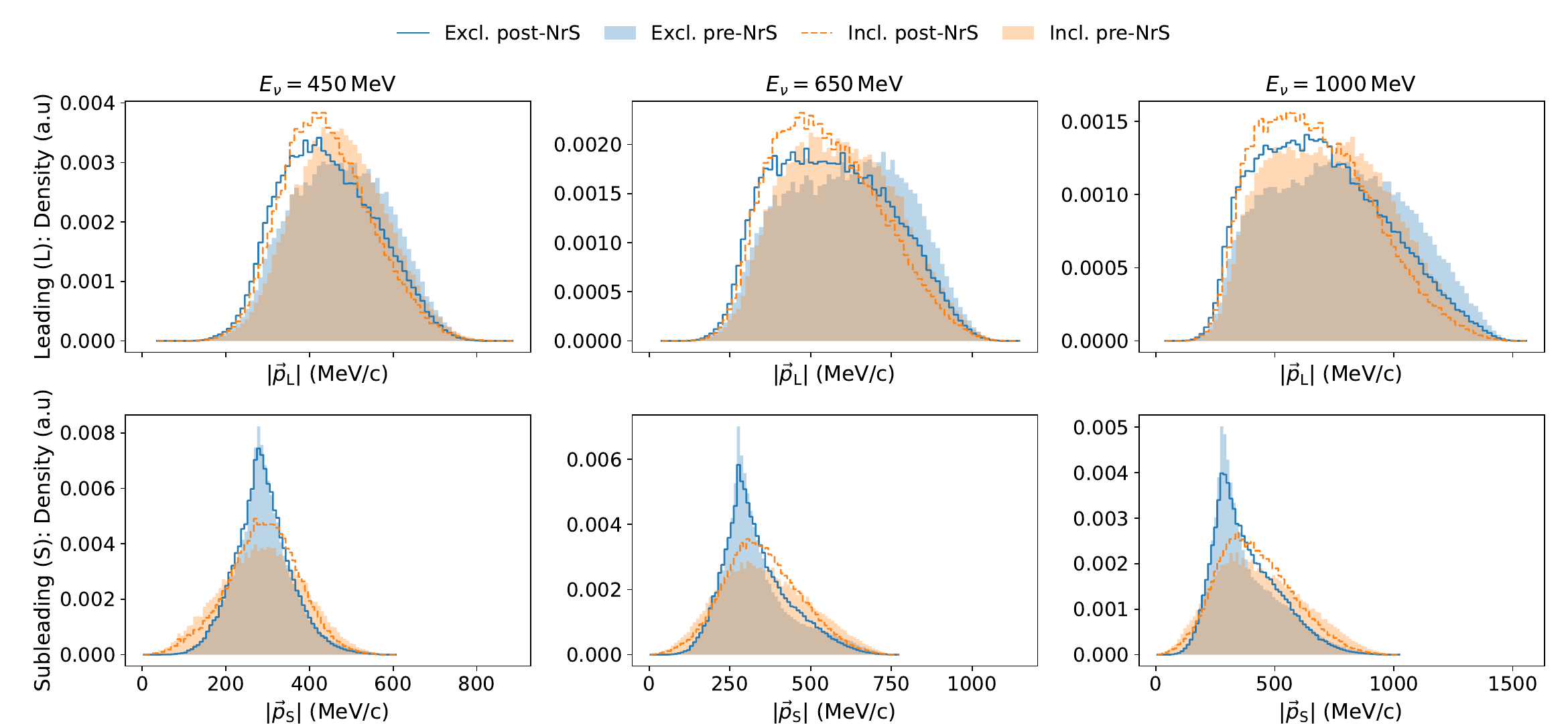}
    \caption{Leading (top) and subleading (bottom) $(pp+pn)$ nucleon momentum distributions (exclusive vs inclusive) before and after NrS $(pp+pn)$.}
    \label{fig:NucleonMomentum_PostFsi}
\end{figure*}
In Fig.~\ref{fig:NucleonMomentum_PostFsi}, we compare the momentum spectra of the leading and subleading nucleons for different neutrino energies. With increasing $E_\nu$, the exclusive prescription yields a harder leading-nucleon spectrum (enhanced high $|\vec{p}\,|$ tail), while the subleading spectrum shows reduced high $|\vec{p}\,|$ strength and is more concentrated at lower momenta compared to the inclusive treatment. The smearing and momentum reduction from NrS can be seen compared to the pre-NrS result. This reflects the exclusive dynamics, in which the nucleon at the primary interaction vertex typically carries a larger share of the transferred four-momentum $q^\mu$.
At the lowest energies the two prescriptions are largely indistinguishable once NrS is applied, but the discrepancies grow with $E_{\nu}$ as higher available momenta accentuate the kinematic differences.

\begin{figure*} [htbp]
    \centering
    \includegraphics[width=\linewidth]{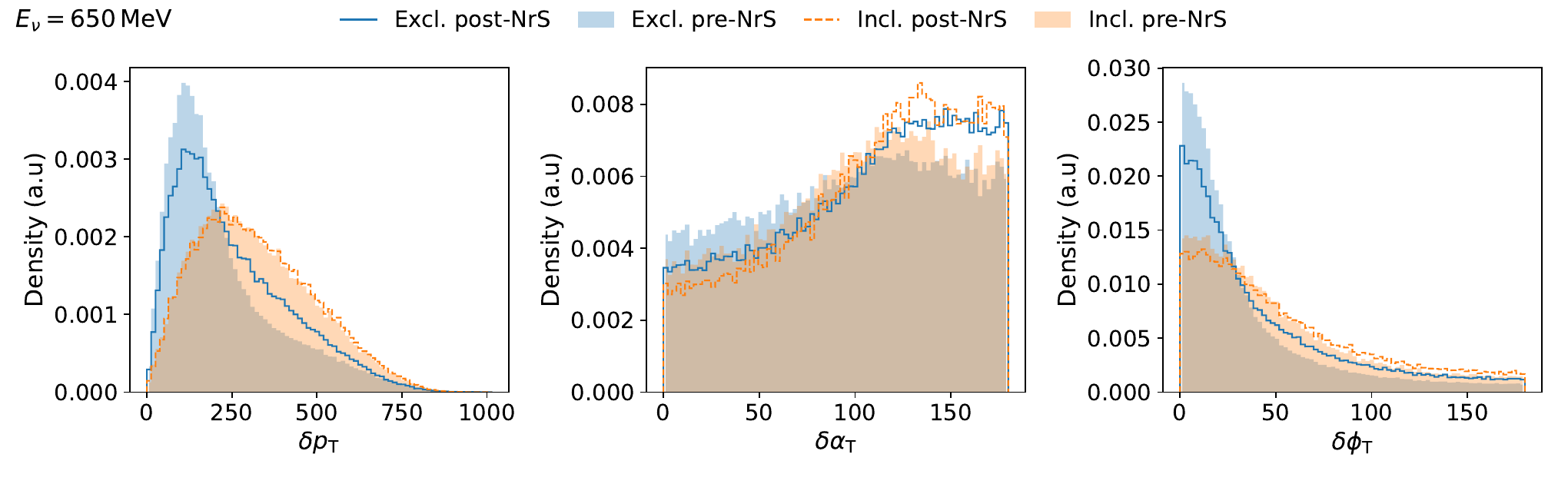}
    \caption{Same as in Fig.~\ref{fig:tKI_Variables_PreFsi}, but with and without considering NrS $(pp+pn)$.}
    \label{fig:tKI_Variables_PostFsi}
\end{figure*}

This behavior is reflected in the TKI distributions shown in Fig.~\ref{fig:tKI_Variables_PostFsi}. 
$\delta p_T$ and $\delta \phi_T$ exhibit minor changes relative to the pre‑NrS results. By contrast, the transverse opening angle $\delta \alpha_T$ is much more effected by NrS, leading to a less apparent difference across both treatments, consistent with its expected sensitivity to rescattering ~\cite{Lu:2015tcr}. Since TKI variables only depend on the leading nucleon, having a preference for a higher momentum leading nucleon in the exclusive treatment effectively leads to a higher probability of surviving NrS effects, while for the inclusive treatment this smearing is spread evenly across the emitted nucleons.

\section{Observability in detectors}
\label{sec:observability}

While the previous sections have focused on the theoretical differences between the inclusive and exclusive kinematic treatments of $2p2h$ interactions, it is crucial to assess the observability of these differences in actual neutrino detectors, especially those with enhanced hadronic detection capabilities. In this section, we explore the potential for current and future neutrino experiments to measure the identified observables and distinguish between the two kinematic treatments.

\subsection{SuperFGD at T2K ND280}
\label{subsec:nd280}

The ND280 near detector at T2K provides the critical constraint on the unoscillated neutrino flux and on neutrino–nucleus interaction systematics used in the oscillation fits. By measuring lepton and hadron kinematics close to the beam, ND280 constrains cross‑section model components and the detector response, reducing systematic uncertainties that would otherwise limit oscillation sensitivity.

The original ND280 design achieved excellent performance for forward muons but had limited acceptance for large‑polar‑angle tracks and for low‑momentum protons; this restricted the experiment's ability to measure exclusive, high‑angle and low‑energy hadronic final states that are particularly sensitive to multinucleon dynamics. The ND280 upgrade addresses these limitations by replacing the central tracker with a high‑granularity three‑dimensional scintillator target (the SuperFGD), adding two high‑angle TPCs and a surrounding time‑of‑flight system \cite{T2K:2019bbb}. These changes yield two essential improvements for the observables studied here: (i) substantially increased muon acceptance (approaching $4\pi$ coverage for tracking), which gives access to larger lepton scattering angles where exclusive $2p2h$ effects are pronounced; and (ii) a dramatically improved hadron detection capability -- the SuperFGD's fine segmentation lowers the effective proton detection threshold and enables tagging of low‑energy protons and delayed signals from neutron‑induced secondaries. Together these upgrades make ND280 far more sensitive to exclusive hadronic final states (leading/subleading nucleons, TKI variables, and nucleon multiplicities) and therefore well suited to test the exclusive vs.\ inclusive $2p2h$ kinematic prescriptions considered in this work \cite{T2K:2019bbb}.

\begin{figure}[htbp]
    \centering
    \includegraphics[width=\linewidth, height=0.9\linewidth]{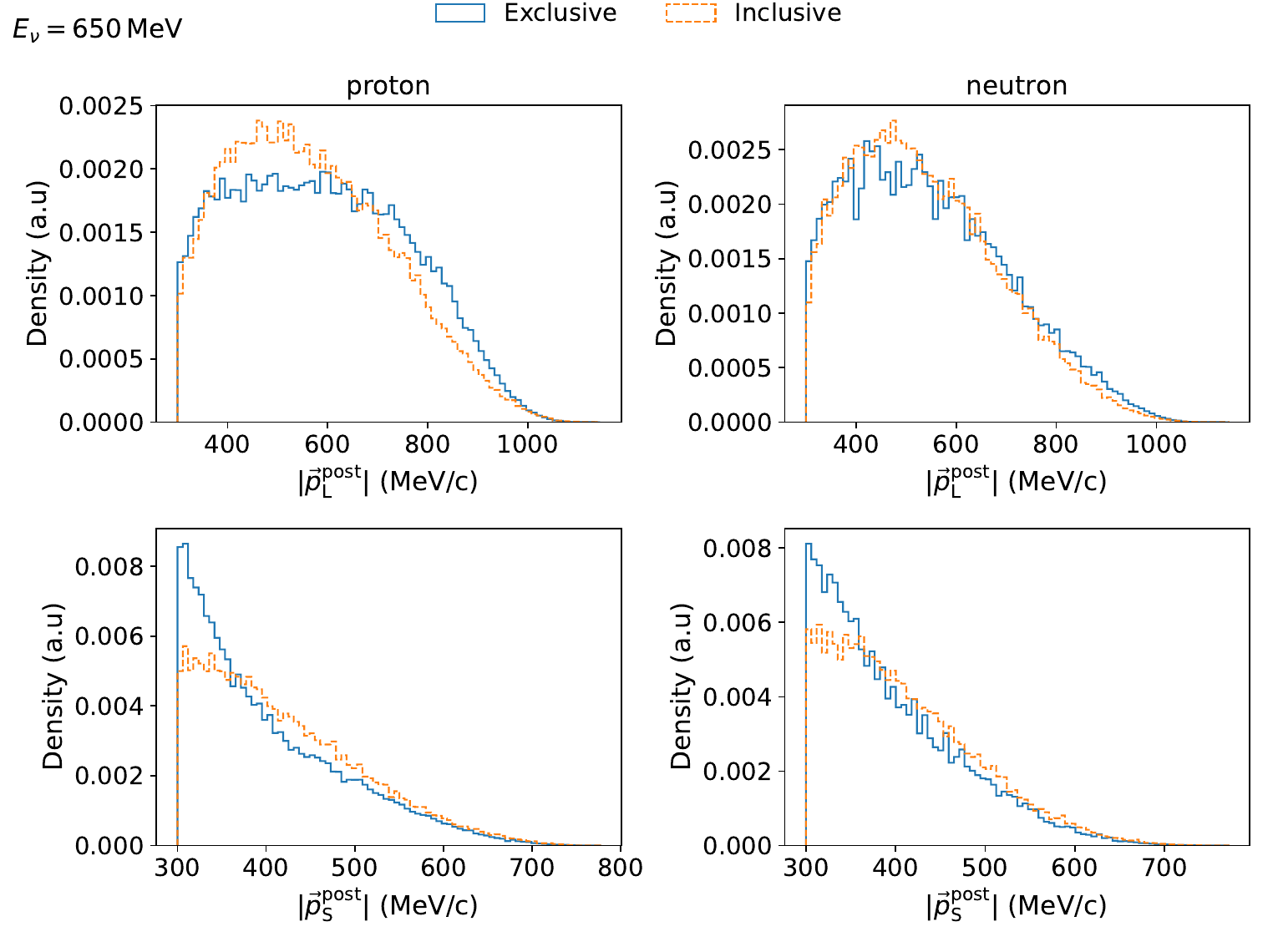}
    \caption{Leading (top) and subleading (bottom) nucleon momentum distribution at $E_\nu = 650$ MeV for $(pp+pn)$ after NrS and ND280 SuperFGD detection threshold cuts for protons (left) and neutrons (right).}
    \label{fig:ND280_momentum_trends_lead_pn}
\end{figure}

\begin{figure*}[h]
    \centering
    \includegraphics[width=\linewidth]{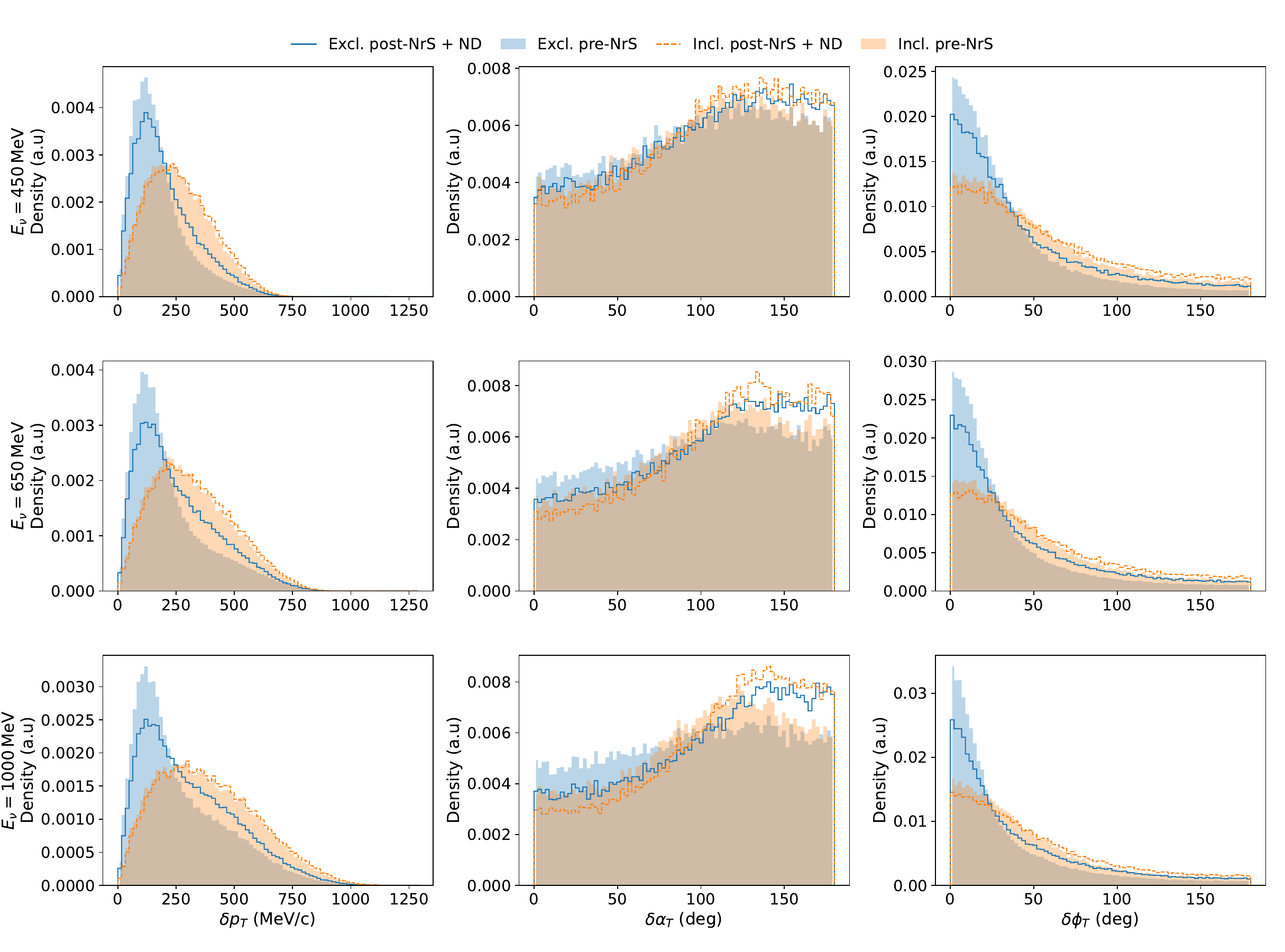}
    \caption{TKI variables (inclusive versus exclusive) $(pp+pn)$ before after NrS and ND280 SuperFGD detection threshold cuts for neutrino energies of 450 MeV (top), 650 MeV (middle) and  1000 MeV (bottom). left, center and right  panels show the obtained distributions for  $\delta p_T$, $\delta \alpha_T$ and $\delta \phi_T$, respectively.}
    \label{fig:ND280_tki_variables}
\end{figure*}

We test the SuperFGD detectability by imposing simple detection thresholds on the final‑state nucleons.  Following preliminary design studies, we apply a proton momentum threshold of $p > 300 \text{ MeV/c}$ and a neutron kinetic‑energy threshold between 45$-$50 MeV (corresponding to roughly 300 MeV/c of momentum). When selecting proton‑only and neutron‑only samples (Fig.~\ref{fig:ND280_momentum_trends_lead_pn}), the impact of the thresholds is most pronounced for protons, consistent with the channel fractions reported in Table~\ref{tab:channel-fractions}.
%
%
These momenta distributions directly impact the TKI variables in Fig.~\ref{fig:ND280_tki_variables}. Here we see that the differences between the inclusive and exclusive treatments are still visible even after the detection thresholds are applied. This both indicates the robustness of these observables to detector effects and highlights the potential for ND280 to discriminate between the two kinematic treatments. It is important to note that these results do not include detector smearing effects, which will further modify the distributions. A full detector simulation study is needed to quantify the sensitivity of ND280 to these differences.

\begin{figure*}[h]
    \centering
    \includegraphics[width=\textwidth]{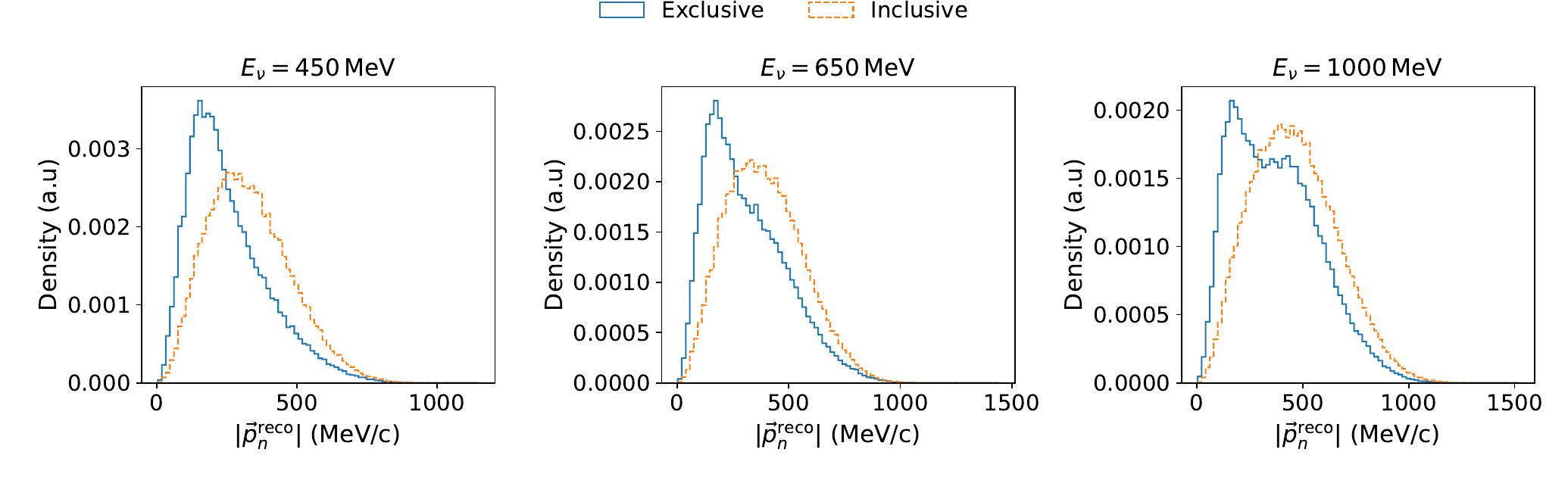}
    \caption{Reconstructed target momentum distributions (see Eq.~\eqref{eq:ptarget})  $(pp+pn)$ after NrS and ND280 SuperFGD detection threshold cuts for $E_\nu = 450$ MeV (left), $650$ MeV (center) and $1000$ MeV (right).}
    \label{fig:ND280_reco_target}
\end{figure*}

The reconstructed target momentum has been shown to be sensitive to multinucleon effects \cite{Filali:2024vpy}, hence having different kinematics can impact these distributions. In Fig.~\ref{fig:ND280_reco_target}, we show the reconstructed target variable defined as:
\begin{equation}
    p_{\rm target}^{\rm reco} = |\vec{p}_\nu - \vec{p}_\mu - \vec{p}_{N_{\rm lead}}| \label{eq:ptarget}
\end{equation}  
where, as introduced above,  $\vec{p}_\nu$ is the incoming neutrino momentum, $\vec{p}_\mu$ is the outgoing muon momentum and $\vec{p}_{N_{\rm lead}}$ is the leading nucleon momentum. This variable is sensitive to the missing momentum in the interaction and can be used to probe multinucleon effects.
We see that the exclusive treatment leads to a narrower distribution compared to the inclusive treatment after applying the detection thresholds. This is expected due to the QE-like behavior of the exclusive treatment leading to a more balanced momentum distribution as compared to the inclusive treatment. It is interesting to note that the differences within the exclusive treatment across neutrino energies leads to a pronounced secondary peak at higher reconstructed target momenta, due to the resonance $\Delta$ contributions. This highlights another important aspect of having access to exclusive kinematics: it gives a more detailed insight into the underlying interaction mechanisms and can help disentangle different contributions to the cross-section without averaging over them, as in the inclusive treatment.


Because the inclusive and exclusive treatments predict different outgoing-nucleon kinematics, they also predict different reconstructed-$E_\nu$ distributions, especially when the subleading nucleon is not detected and the reconstruction uses only the lepton and leading nucleon. This can bias oscillation and cross-section measurements that rely on reconstructed neutrino energy.

\section{Conclusions}
\label{sec:concl}

In this work we have quantified the differences between the inclusive and exclusive kinematic treatments of $2p2h$ interactions in neutrino–nucleus scattering.
At the nucleon level we identified observables that are particularly sensitive to the microscopic kinematics: leading and subleading nucleon momenta, detailed lepton–hadron correlations, and transverse kinematic imbalance  variables. 
The exclusive calculation produces a markedly more asymmetric sharing of energy and momentum between the two outgoing nucleons; one nucleon frequently carries a dominant fraction of the transferred four‑momentum and this feature survives realistic intranuclear cascade smearing.

The upgraded ND280 (SuperFGD) and future high‑resolution trackers offer genuine potential to test these model differences. 
Even after applying plausible detection thresholds,  the exclusive prescription leaves visible signatures in the leading‑nucleon spectra and in TKI distributions; with full detector simulation and unfolding these signatures can be turned into experimental constraints. 
To realise this potential we recommend a program combining (i) full detector‑level studies (including smearing and efficiencies), (ii) targeted experimental selections optimized for high lepton polar angle and multi‑nucleon final states, and (iii) joint lepton+hadron fits that exploit complementary observables (proton/neutron samples, multiplicities, TKI) to maximise sensitivity to exclusive dynamics.

A technical but crucial point for practical implementation is that the exclusive Valencia calculation cannot be represented economically by the low‑dimensional hadron‑tensor tables commonly used for inclusive models. 
Exclusive kinematics require retaining initial nucleon momenta and multi‑particle phase‑space information, which leads to a prohibitively large multidimensional response if tabulated naively. 
Addressing this challenge requires novel approaches such as on‑the‑fly exclusive samplers or fast surrogate models (e.g. normalizing flow emulators) \cite{ElBaz:2025qjp}. 
Moreover, full exploitation of these exclusive predictions will require integration into mainstream event generators and comprehensive detector simulations to quantify discriminating power—efforts that demand dedicated methodological development and are the subject of ongoing and future publications.

Taken together, these developments position the exclusive Valencia framework as both a tool for improved theory–data comparisons and a pathway to exploit forthcoming detector capabilities and reduce modeling uncertainties in oscillation measurements.

\begin{acknowledgments}
We sincerely thank N. Rocco for her valuable comments.
This work has been partially supported by the Spanish 
MICIU/AEI/10.13039/501100011033 under 
grants   PID2023-147458NB-C21 and CEX2023-001292-S, by 
Generalitat Valenciana's PROMETEO
grant CIPROM/2023/59, by the “Planes Complementarios de I+D+i” program
(Grant No. ASFAE/2022/022) from MICINN with funding from
the European Union NextGenerationEU and Generalitat
Valenciana. We gratefully acknowledge the Swiss National Science Foundation (SNSF) for financial support under grant number 200020\_204609.
\end{acknowledgments}

\appendix







\bibliographystyle{apsrev4-2}
\bibliography{apssamp}

\end{document}